\documentclass[12pt]{article}
\usepackage{amsmath,euscript,array,amssymb,cite}
\newcommand{\PSbox}[3]{\mbox{\rule{0in}{#3}\includegraphics{#1}\hspace{#2}}
}
\newcommand{\beq}{\begin{eqnarray}}
\newcommand{\eeq}{\end{eqnarray}}
\newcommand{\drawsquare}[2]{\hbox{%
\rule{#2pt}{#1pt}\hskip-#2pt
\rule{#1pt}{#2pt}\hskip-#1pt
\rule[#1pt]{#1pt}{#2pt}}\rule[#1pt]{#2pt}{#2pt}\hskip-#2pt
\rule{#2pt}{#1pt}}

\newcommand{\Yfund}{\raisebox{-.5pt}{\drawsquare{6.5}{0.4}}}

\newcommand{ \sla }[1]{\setbox0=\hbox{$#1$}         
    \dimen0=\wd0                                     
    \setbox1=\hbox{/} \dimen1=\wd1                   
    \ifdim\dimen0>\dimen1                            
       \rlap{\hbox to \dimen0{\hfil/\hfil}}          
       #1                                            
    \else                                            
       \rlap{\hbox to \dimen1{\hfil$#1$\hfil}}       
       /                                             
    \fi}                                             %

\makeatletter
\def\vereq#1#2{\lower3pt\vbox{\baselineskip1.5pt \lineskip1.5pt
\ialign{$\m@th#1\hfill##\hfil$\crcr#2\crcr\sim\crcr}}}
\makeatother

\begin{document}
\begin{titlepage}
\begin{center}
     \hfill {\tt hep-th/0110188}\\
\vskip 0.2in

{\LARGE \bf Exact Results in 5D  from  Instantons  and\\ \vskip 0.05in
   Deconstruction}

\vskip 0.25in
{\bf Csaba Cs\'aki$^a$, Joshua
Erlich$^a$, Valentin V.~Khoze$^b$, \\
Erich Poppitz$^c$, Yael Shadmi$^d$ and  Yuri Shirman$^e$}

\vskip 0.2in

$^a${\em Theoretical Division T-8, Los Alamos National Laboratory,
Los Alamos, NM 87545, USA}
\vskip 0.1in
$^b${\em Department of Physics, University of Durham,
Durham, DH13LE, UK}
\vskip 0.1in
$^c${\em Department of Physics,
University of Toronto, Toronto, Ontario, M5S1A7, Canada}
\vskip 0.1in
$^d${\em Department of Physics, Technion, Technion City,
32000 Haifa, Israel}
\vskip 0.1in
$^e${\em Department of Physics,
California Institute of Technology, Pasadena, CA 91125, USA}
\vskip 0.1in

{\tt  csaki@lanl.gov, erlich@lanl.gov, valya.khoze@durham.ac.uk,
poppitz@physics.utoronto.ca, yshadmi@physics.technion.ac.il,
yuri@theory.caltech.edu}
\end{center}

\vskip .5in
\begin{abstract}
We consider non-perturbative effects in theories with extra dimensions
and the deconstructed versions of these theories. We establish the rules 
for
instanton calculations in 5D theories on the circle, and use them for
an explicit one-instanton calculation in a supersymmetric gauge theory. 
The
results are then compared to the known exact Seiberg-Witten type solution
for this theory, confirming the validity both of the exact results
and of the rules for instanton calculus for extra dimensions introduced
here.
Next we consider the non-perturbative results from the perspective of
deconstructed extra dimensions. We show that the
non-perturbative results of the deconstructed theory do indeed reproduce
the known results for the continuum extra dimensional theory, thus
providing
the first non-perturbative evidence in favor of deconstruction.
This way deconstruction also allows us to make exact predictions in 
higher
dimensional
theories which agree with earlier results,
and helps to clarify the interpretation of 5D instantons.

\end{abstract}
\end{titlepage}

\newpage

\section{Introduction}
\setcounter{equation}{0}
\setcounter{footnote}{0}
Theories with extra dimensions might play an important role in 
resolving a
variety of outstanding issues in particle physics: they might
resolve the hierarchy problem \cite{hierarchy}, give new mechanisms
for communicating supersymmetry breaking \cite{susybreaking},
or yield new insights into the flavor problem and proton stability
\cite{AS}. In many of the interesting applications \cite{susybreaking,AS}
the gauge sector
of the SM propagates in the extra dimension (though not in the
models of \cite{hierarchy} which aim to solve the hierarchy
problem).
If the gauge fields do propagate along the extra dimension, then
non-perturbative effects in the low-energy effective theory may
differ significantly from those in ordinary 4D theories. The reason is
that once the extra dimension is compactified, the instanton can wrap
the compact extra dimension. Therefore,  the
presence of the extra dimension itself will modify the rules for
instanton calculus and   influence the resulting non-perturbative 
effects.

In this paper,  we initiate
the study of non-perturbative effects
for extra dimensional model building, using
explicit instanton calculations, existing exact results in higher
dimensional
gauge theories
\cite{Nekrasov,Seiberg:1996bd,Intriligator:1997pq,Witten:1996qb,Marshakov,
Kol,LN,higherd},
and deconstruction
\cite{deconstruction}. We will concentrate on a single
extra dimension compactified on a circle. In 5D with all dimensions
non-compact there are no known finite action instanton configurations
that would contribute to the semi-classical expression for the
path integral. Ordinary 4D instantons would give
a diverging action once integrated over the fifth coordinate (assuming
that the 4D instanton is independent of the fifth coordinate), and
no fully localized 5D instanton solutions are known to exist.
This situation  changes drastically once the fifth coordinate
is compactified. In this case the ordinary 4D instanton does give a
finite contribution. In addition there is a tower of instantons
that contribute, due to the fact that the 4D instanton can wrap the extra
dimension. In order to gain control over the non-perturbative
effects in a strongly interacting theory we will be considering
supersymmetric extra dimensional theories.
The simplest such theory is an $SU(2)$ gauge theory with 8
supercharges in 5D (which corresponds to
${\cal N}=2$ supersymmetry in 4D). The reason behind the doubling
of the minimal number of supercharges is that in 5D the Dirac spinor
is irreducible. The aim of considering
this model is not to build a realistic theory with extra dimensions,
but rather to establish the rules for instanton calculations in the
presence of extra dimensions, which can later be applied to more
realistic models. Since in this toy model the effective 4D theory
is an ${\cal N}=2$ theory, it can be exactly solved in terms of a
Seiberg-Witten curve \cite{SW1,SW2}. This solution was first
proposed by Nekrasov in \cite{Nekrasov}.

We begin the first part of this paper by
reviewing Nekrasov's solution, and slightly modify it to
account for an ambiguity in the Seiberg-Witten curve.
This ambiguity is analogous to those appearing in the
ordinary 4D Seiberg-Witten results discussed in \cite{MO2}.
We then turn to an explicit instanton
calculation to verify the exact results of the curve. During the course
of this calculation we show that  there are two towers of instantons that
contribute to the effective action. One of these towers
is comprised by the large gauge transformed versions of the ordinary
4D instanton
wrapping the extra dimension $n$ times. The second tower is obtained by
applying an ``improper'' gauge transformation on the instanton solution,
and the corresponding large gauge transformed versions of the solution
obtained this way. This improper gauge transformation is not among the
allowed gauge transformations of the theory, since it does not obey
the necessary boundary condition. Nevertheless, the transformed
instanton solution obtained this way does obey all the
conditions for a proper semiclassical solution. A sum over
these two towers of instantons does indeed reproduce the
exact results.
Thus, our calculation confirms and improves the exact results,
and more importantly it establishes the rules
for instanton calculus in 5D theories.
The agreement of the two calculations confirms that
there are no fully localized 5D instantons, and that the full
semiclassical results can be obtained by the sum over the two instanton
towers.

A recent major development in the field of extra dimensions is the
construction of 4D gauge theories which reproduce the effects of
extra dimensions. The ``deconstructed'' theory
\cite{deconstruction} is based
on a product gauge group theory in 4D,
and in fact provides a latticized version of
the extra dimensional theory. This has several interesting applications
for model building in four dimensions
\cite{ACG2,Fermi2,CEGK,CKSS,CKT,otherdec}.
So far, the equivalence between the
deconstructed theory and the higher dimensional models has been purely
based on perturbative arguments, like matching of the perturbative mass
spectra of the two theories. In the second part of this paper, we provide
the
first evidence that deconstruction captures
the non-perturbative effects as well. Deconstruction of the simplest
5D supersymmetric theory was done in \cite{CEGK}.\footnote{Very
recently deconstruction of 6D supersymmetric theories has been considered
in \cite{littlestring}.}
The deconstructed version of the theory turns out to be
the ${\cal N}=1$, 4D product group theory considered in \cite{CEFS}, 
where
some non-perturbative results for this theory were obtained.
Since the deconstructed theory only has
${\cal N}=1$ supersymmetry, one cannot provide a full solution to the
low-energy effective theory, like the Seiberg-Witten solutions;
exact results are restricted to the holomorphic quantities in the 
theory---in
this case, the gauge kinetic term which includes the gauge coupling. We 
will
show that the non-perturbative information that can be easily extracted
from
the deconstructed theory agrees with results from the continuum
theory. This then serves partly as an independent derivation
of the non-perturbative results for the 5D theory, which have previously
been obtained from symmetry and consistency requirements, and
also shows that the deconstructed theory does indeed capture the
non-perturbative effects of the higher dimensional theory.

This paper is organized in two major parts: Section 2, devoted to an 
analysis
of the 5D theory on the circle
and its low-energy nonperturbative dynamics, and Section 3,
containing the corresponding analysis of
the deconstructed theory and
a comparison with the compactified continuous theory.

We begin, in Section 2.1, with a review of the 4D Seiberg-Witten setup  
and of
the curve describing the low-energy dynamics of the 5D theory on the 
circle
due to Nekrasov (Section 2.2).
In Section 2.3 we derive the rules for instanton calculations in the 
compactified
supersymmetric 5D theory.
We  show that a summation over two infinite towers of instantons is 
required to
restore invariance
under the proper and improper large gauge transformations.
We  perform a  one-instanton calculation of the contribution to the 
low-energy
$\tau$ parameter of the theory and
show that the result is in agreement with the improved Nekrasov curve.

We begin the study of the dynamics of the deconstructed theory in 
Section 3.
We review  the deconstructed theory and its Seiberg-Witten  curve  in 
Section 3.1.
The matching of the perturbative mass spectra between the deconstructed 
and
continuous 5D theories is reviewed in Section 3.2.  After that, in 
Section 3.3,
we study  the
correspondence between continuum and deconstructed instantons. We show, 
using the
brane picture of the
deconstructed theory, how the proper and improper large gauge 
transformations arise
in deconstruction, and
argue that the contribution of the diagonal $(1,1,1,...., 1)$-instantons 
of the
deconstructed theory match those
of the two infinite towers of instantons of the continuum theory.
In Section 3.4 we derive the continuum Seiberg-Witten curve
from the deconstructed theory and show that it matches the curve of the 
continuous
theory.
Section 3.5 is devoted to a detailed discussion of the matching of  
moduli between
the continuous and deconstructed theories. This is an important issue,
   somewhat complicated by the fact that relations between moduli
receive  corrections from the quantum modification of the moduli space  
of the
individual gauge groups of the deconstructed theory.
  In Section 3.5.1, we give a heuristic argument in favor of the correct 
matching.
We strengthen this argument by an explicit instanton calculation 
(Section 3.5.2)
showing that the  modulus, which is to be identified with the continuum 
theory modulus,
does not receive corrections from  instantons in the broken gauge groups.
  In Section 3.5.3, motivated by the brane picture,  we point out
  the existence of a special flat direction where corrections
to the
holomorphic deconstructed theory moduli from instantons in
the broken gauge groups vanish.
Finally, in Section 3.6, we show that the large radius limit of the 
low-energy
$\tau$ parameter  has the behaviour required by  5D nonrenormalization 
theorems.

\section{5D $SU(2)$ Curve and Explicit Instanton Calculations}
\setcounter{equation}{0}
\setcounter{footnote}{0}

In this Section, we first review the solution of the 5D, ${\cal N}=2$ 
pure $SU(2)$
gauge theory, in terms of a Seiberg-Witten type curve,
and then show how to perform an explicit
instanton calculation in the theory. We will explain how to obtain
the relevant instanton contributions from the ordinary 4D instanton, and
find that the result of the explicit calculation agrees with the curve 
prediction.

We will    perform
  a 5D calculation of the path-integral contributions of 4D instantons,
summing  over two infinite towers of instanton solutions.  Every solution
we sum over
is obtained  as an $x_5$-dependent large gauge transformation
of the usual 4D $x_5$-independent instanton, giving it a nontrivial
dependence on the compactified coordinate.
These instantons have precisely
the same number of bosonic and fermionic zero modes as the conventional
4D instantons.
In addition, due to  supersymmetry and self-duality of
the instanton background, all contributions of non-zero modes to the
determinants and higher loops in the instanton
background cancel, as in the 4D case. The  dependence of
the instanton amplitude on the instanton size
   is determined entirely
by the number of bosonic and fermionic zero modes and  is
the same as in the 4D instanton calculation (in particular,
   the dependence on the instanton size in the compactified 
supersymmetric theory is
controlled by the 4D ${\cal N}=2$ beta function).  Thus the instanton effects 
in the supersymmetric
5D theory turn out to be renormalizable in the 4D sense.  Therefore, it 
is
meaningful to compare instanton-induced nonperturbative effects   in the
continuum 5D and  deconstructed large-$N$ 4D theories.

\subsection{4D Seiberg-Witten set-up}

First, let us introduce standard notation for the ordinary Seiberg-Witten
case
\cite{SW1}.
Consider pure ${\cal N}=2$ SU(2) theory in 4D. On the Coulomb branch
the adjoint scalar field of the ${\cal N}=2$ vector superfield develops a
vev
\begin{equation}
\langle \phi \rangle = a {\sigma^3 \over 2} \ ,
\end{equation}
and the gauge-invariant modulus $u$ is defined via:
\begin{equation}
u=\langle {\rm Tr} \phi^2 \rangle \ .
\end{equation}
In the weak coupling regime $u$ is given by
\begin{equation}
u= {a^2\over 2} + \sum_{k=1}^\infty G_k {\Lambda^{4k} \over a^{4k-2}}
  \ , \label{uexp}
\end{equation}
where the infinite sum represents instanton contributions. Here $k$ is
the instanton number and $\Lambda$ is the dynamical scale of the theory.
The complexified coupling $\tau_{\rm SW}$ is given by the second 
derivative
of the
holomorphic prepotential ${\cal F}_{\rm SW}$
\begin{equation}
\tau_{\rm SW} (a)= {\partial^2 {\cal F}_{\rm SW}(a) \over \partial a^2}=
{4\pi i \over g^2(a)} + {\vartheta(a) \over 2 \pi}
  \ .\label{tausw}
\end{equation}
In the weak coupling regime it receives contributions in perturbation
theory at one loop and from all orders in instantons
\begin{equation}
\tau_{\rm SW}(a)= {i\over \pi} \log{a^2\over \Lambda^2} +
\sum_{k=1}^\infty \tau_k {\Lambda^{4k} \over a^{4k}}
  \ . \label{tauexp}
\end{equation}
The low-energy dynamics of the theory
can be determined from a genus one auxiliary Riemann surface described by
an elliptic Seiberg-Witten curve. The curve is given by
\begin{equation}
y^2=(x^2-\Lambda^4)(x-u) \ ,
\label{swcone}
\end{equation}
where $x$ and $y$ parameterize the surface. The first step toward 
obtaining
the exact low-energy effective action for the Seiberg-Witten theory is to
define the meromorphic differential $\lambda$
\begin{equation}
\lambda={y \,dx\over \Lambda^4-x^2} \ .
\label{swdif}
\end{equation}
Then the vevs of the scalar, $a,$ and of the dual scalar, $a_D,$
are determined as functions of the modulus $u$ by integrating
the meromorphic form $\lambda$ over the appropriately chosen cycles
$\gamma_a$ and $\gamma_{a_D}$ of the
Riemann surface \eqref{swcone}:
\begin{equation}
a(u)= {\sqrt{2} \over 2 \pi} \oint_{\gamma_a} \lambda =
{\sqrt{2} \over \pi} \int_{-\Lambda^2}^{\Lambda^2}\lambda \ ,
\end{equation}
\begin{equation}
a_D(u)\equiv{\partial {\cal F}(a) \over \partial a}
= {\sqrt{2} \over 2 \pi} \oint_{\gamma_{a_D}} \lambda =
{\sqrt{2} \over \pi} \int_{\Lambda^2}^{u}\lambda \ .
\end{equation}
Combining these expressions one can obtain ${\cal F}(a)$, which in turn
determines the complete low-energy effective action for an ${\cal N}=2$
theory.
A few comments are in order. First, the dynamical scale of the theory
is defined in the so-called Seiberg-Witten scheme. It is related
\cite{FP} to the
Pauli-Villars (PV) or $\overline{{\rm DR}}$ scheme (which are used for
explicit perturbative
and instanton calculations) via the one-loop exact expression,
$\Lambda^2 =2\Lambda_{\rm PV}^2=2\Lambda_{\rm
\overline{DR}}^2.$\footnote{In
this Section we will use the Seiberg-Witten scheme, while in Section 3,
we use the  $\overline{DR}$  scales. This difference will only be 
important
for our comparison of $\tau$ parameters and is trivial to account for.}
The integrals in the expressions for $a(u)$ and $a_D(u)$ can be easily
evaluated.
In particular, in the weak-coupling regime, $a\gg \Lambda$, the 
expression
for $a(u)$ can be inverted giving the modulus \eqref{uexp},
and then the expression for
$a_D(u(a))$ can be differentiated with respect to $a$ to determine the
coupling \eqref{tauexp}. All the coefficients of these expansions can be
obtained from the exact solution above. In particular, in the
Seiberg-Witten
scheme the one-instanton coefficients are
\begin{equation}
G_1={1\over 4} \ , \qquad \tau_1=-{i\over \pi}{3\over 4} \ .
\label{uoto}
\end{equation}
In fact, for all instanton numbers the instanton contributions to
$\tau_{\rm SW}$ and $u$ are related via the
Matone relation \cite{Matone, DKMmat}
\begin{equation}
\tau_k\,=\,-{i\over \pi}\,{(4k-2)(4k-1)\over 2k}\, G_k \ .
\label{matoner}
\end{equation}
Alternatively, these coefficients for $k=1,2$
can be derived \cite{MO1} via direct multi-instanton
calculation of the effective action.

Now, following Nekrasov \cite{Nekrasov}
(and keeping all the numerical factors in place)
  we make a change of variables:
\begin{eqnarray}
y&=&i {p \over \sqrt{2}} ~\Lambda^2 \sinh(q) \ , \\
x&=&\Lambda^2 \cosh(q) \ .
\end{eqnarray}
The Seiberg-Witten curve becomes
\begin{equation}
u={p^2 \over 2}+\Lambda^2 \cosh(q) \ ,
\label{swct}
\end{equation}
and the meromorphic differential is now
\begin{equation}
\lambda=-{i\over \sqrt{2}}\,pdq \ .
\label{merdt}
\end{equation}
The vevs $a(u)$ and $a_D(u)$ are given by
\begin{equation}
\vec{a}(u)=(a_D(u),a(u))= -{i\over 2\pi}\oint_{\vec{\gamma}} pdq \ .
\label{periods}
\end{equation}
The cycles $\vec{\gamma}$ are chosen in such a way that the correct
asymptotic behaviour of $a(u)$ and $a_D(u)$ as $u\to \infty$ is 
reproduced.

In particular we have
\begin{equation}
a(u)= -{i\over 2\pi}\int_{-i\pi}^{i\pi} pdq \rightarrow \sqrt{2u} \ ,
\label{auper}
\end{equation}
in agreement with \eqref{uexp}.
For future use we introduce two new parameters,
\begin{equation}
w\equiv \sqrt{2u} \ , \qquad
\nu_4\equiv {\Lambda^2 \over u} \ ,
\label{defswn}
\end{equation}
and rewrite \eqref{periods} in the convenient form:
\begin{equation}
{\partial \vec{a}(u)\over \partial w}
= -{i\over 2\pi}\oint_{\vec{\gamma}}
{dq \over \sqrt{1-\nu_4\cosh(q)}} \ .
\label{pertt}
\end{equation}

\subsection{The improved 5D $SU(2)$ Seiberg-Witten curve}

The ${\cal N}=1$
5D SU(2) theory on ${\bf R}^4\times S^1$ will be viewed from
the perspective of the low-energy effective 4D theory,
i.e. all the 5D fields are represented as infinite sets
of KK-modes which are functions of the ${\bf R}^4$-variables.

There is a complex scalar $\Phi= \phi + i A_5$, which develops the
vev
\begin{equation}
\langle \Phi \rangle =  A {\sigma^3 \over 2} \ ,
\end{equation}
and the gauge-invariant modulus $U$ is now defined as \cite{Nekrasov}
\begin{equation}
U=  {1\over 2}~\left\langle {\rm Tr}{\cosh(2\pi\Phi R) \over \pi^2 R^2}
\right\rangle \ ,
\label{eq:U}
\end{equation}
which has the weak-coupling expansion
\begin{equation}
U={\cosh(\pi A R) \over \pi^2 R^2} + {\rm instantons} \ .
\label{Uexp}
\end{equation}

We claim that
the curve describing the low-energy dynamics of the theory is given by
\begin{equation}
U=\,{1\over \pi^2 R^2}\,\cosh(\pi R p)\,\sqrt{1+2
(\pi R\Lambda)^2\,f(\pi R\Lambda)\, \cosh(q)}
\ . \label{curve5d}
\end{equation}
Here $\Lambda$ is exactly the same dynamical scale as before in
\eqref{swcone}.
Notice that the curve \eqref{curve5d}
is slightly different from Nekrasov's relativistic generalization
of Toda's chain \cite{Nekrasov}: The expression on the right hand side of
\eqref{curve5d} contains an a priori unknown function $f(\pi R\Lambda)$,
which can not be determined based on symmetry arguments only.
This function is just 1 classically, but it can get instanton corrections
at every level $k$:
\begin{equation}
f(\pi R\Lambda)= 1+ \sum_{k=1}^\infty f_k (\pi R\Lambda)^{4k}
\ , \label{funf5d}
\end{equation}
where each coefficient $f_k$ has to be determined from an explicit
$k$-instanton
calculation. We will see below that
the function $f$ will be needed to remove certain constant
(i.e. vev-independent) contributions from the complexified coupling
$\tau(A)$.
The ambiguity in the curve predictions introduced by $f$
is similar in spirit to the ambiguities \cite{MO2}
of the Seiberg-Witten curves in the presence of matter.

In terms of this curve,
the vevs $A(U)$ and $A_D(U)$ are determined in exactly the same way as in
\eqref{periods}
\begin{equation}
\vec{A}(U)=(A_D(U),A(U))= -{i\over 2\pi}\oint_{\vec{\gamma}} pdq \ .
\label{pertwo}
\end{equation}
The cycles $\vec{\gamma}$ are the same as in \eqref{periods}
such that
\begin{equation}
A(U)= -{i\over 2\pi}\int_{-i\pi}^{i\pi} pdq \rightarrow {1\over \pi R}
\cosh^{-1}(\pi^2 R^2 U)
  \ ,
\label{aupertwo}
\end{equation}
in agreement with \eqref{Uexp}.
In terms of the new parameters,
\begin{equation}
\cosh(\alpha)\equiv\, \pi^2 R^2 U \ , \qquad
\nu_5 \equiv\, {2f(\pi R\Lambda)\,(\pi R\Lambda)^2\over \sinh^2(\alpha)}
  \ ,
\label{defsan}
\end{equation}
equations \eqref{pertwo} can be expressed
\cite{Nekrasov} in the form of \eqref{pertt}
\begin{equation}
{\partial \vec{A}(U)\over \partial \alpha}
= -{i\over 2\pi}\oint_{\vec{\gamma}} {1\over \pi R}
{dq \over \sqrt{1-\nu_5\cosh(q)}} \ .
\label{perthree}
\end{equation}
Hence, when $\nu_4=\nu_5$, i.e.
\begin{equation}
u=\tilde{U} \equiv {\Lambda^2 \over 2f(\pi R\Lambda)\,(\pi
R\Lambda)^2}(\pi^4 R^4 U^2-1)
\ ,
\label{Utdef}
\end{equation}
the vevs of the two theories are simply related to each other,
\begin{equation}
{\partial \vec{A}\over \partial \alpha}
=\, {1\over \pi R}\, {\partial \vec{a}\over \partial
w}\bigg\vert_{u=\tilde{U}}
\ . \label{form1}
\end{equation}
{}From this we can instantly calculate $\tau$ as a function of the
modulus $U$ of the 5D theory,
\begin{equation}
\tau(U)=
{\partial A_D\over \partial A}(U)
={\partial a_D\over \partial a}(u=\tilde{U})=\tau_{\rm 
SW}(u=\tilde{U})  ~.
  \label{tauform}
\end{equation}
Here on the left hand side we have the coupling
$\tau$ of the 5D theory and on the right
hand side we have the coupling
  $\tau_{\rm SW}$ of the 4D Seiberg-Witten theory.

{}From \eqref{tauform},\eqref{form1} and \eqref{tauexp}
it is easy to determine the 5D coupling at one-loop order,
\begin{equation}
\tau^{\rm pert}(U)=
  {i\over \pi} \log \left({\sinh^2(\pi A R)\over \pi^2 
R^2\Lambda^2}\right)
\label{taupert}
\end{equation}
We will discuss the interpretation of the  perturbative part of $\tau$ in
Section \ref{largeradius}.

Now we will determine $\tau(A)$ in the 5D theory at the 1-instanton 
level.
In order to do this we will
\begin{itemize}
\item determine $A=A(U)$ from \eqref{form1},
\item invert it as $U=U(A)$
and express it as $\tilde{U}=\tilde{U}(A)$ using \eqref{Utdef},
\item calculate $a^2(A)$ via $a^2=a^2(u=\tilde{U}(A))$,
\item finally obtain $\tau(A)=\tau_{\rm SW}(a^2(A))$.
\end{itemize}
Each of these steps is explained in detail below:

{\flushleft {\bf 1.}} At the 1-instanton level equation \eqref{uexp} can 
be
inverted,
\begin{equation}
a^2=2u - G_1{\Lambda^{4} \over u}
  \ . \label{uexp1i}
\end{equation}
Substituting this to the right hand side of \eqref{form1} we get
\begin{equation}
{\partial A\over \partial \alpha}
=\, {1\over \pi R}\, {\partial a\over \partial w}|_{u=\tilde{U}}
=\, {1\over \pi R} \left(1+ {3G_1(\Lambda \pi R)^4 \over
\sinh^4(\alpha)}\right)
\ . \label{form12}
\end{equation}
Integrating this with respect to $\alpha$ we obtain
\begin{equation}
\pi R A = \alpha + G_1\,(\Lambda \pi R)^4 \, {\cosh(\alpha) \over
\sinh(\alpha)}
\left(2-{1\over \sinh^2(\alpha)}\right)
\ . \label{form13}
\end{equation}

{\flushleft {\bf 2.}} Evaluating $\cosh$ of both sides of \eqref{form13}
and using
the definition of $\alpha$ \eqref{defsan},
\begin{equation}
U = {\cosh(\pi R A)\over \pi^2 R^2}\left(1-
G_1(\Lambda \pi R)^4
\left(2-{1\over \sinh^2(\pi R A)}\right)\right)
\ . \label{form14}
\end{equation}
By (\ref{Utdef}) we then determine $\tilde{U}(A)$,
\begin{equation}
\tilde{U} = {\sinh^2(\pi R A)\over 2 \pi^2 R^2}\left(1-(\Lambda \pi
R)^4\left(
f_1 +4G_1 {\cosh^2(\pi R A)\over \sinh^2(\pi R A)}
-2G_1 {\cosh^2(\pi R A)\over \sinh^4(\pi R A)}\right)\right)
\ . \label{form15}
\end{equation}
In deriving the last expression we used the definition of
  $f$, \eqref{funf5d}, in the 1-instanton approximation.

{\flushleft {\bf 3.}}  From \eqref{uexp1i} we determine $a^2(A)$ as
\begin{equation}
a^2=2\tilde{U}(A) - G_1{\Lambda^{4} \over \tilde{U}(A)}
  \ . \label{uexp2i}
\end{equation}

{\flushleft {\bf 4.}}  Finally, we can write down the expression for
$\tau(A)=\tau_{\rm SW}(a^2(A))$ via (\ref{tauexp}),
\begin{equation}
\tau= {i\over \pi} \log \left({\sinh^2(\pi A R)\over \pi^2
R^2\Lambda^2}\right)
- (\Lambda \pi R)^4\left(
f_1 {i\over \pi} + 4G_1 {i\over \pi} +
2 G_1 {i\over \pi}{1\over\sinh^2(\pi A R)} + \tau_1 {1\over\sinh^4(\pi A
R)}
\right)
  \ . \label{tau1ic}
\end{equation}
This expression together with \eqref{uoto} constitutes the 
curve-prediction
for
the coupling of the 5D theory. Now we will verify this prediction
with an explicit 1-instanton calculation. As a by-product of this
comparison
we will also determine $f_1=-4G_1.$

\subsection{Rules for instanton calculations and results}

In order to carry out the explicit instanton calculation we first
need to identify the classical instanton solutions in this theory.
As mentioned before, there are no known instantons in a 5D theory
with all dimensions infinitely large; that is, there are no fully 
localized
5D
instanton solutions. Once one of the dimensions is compactified,
the action of an ordinary 4D instanton (which is assumed to be 
independent
of the coordinate of the extra dimension) will become finite. However,
it turns out that this is not the only finite action solution that
exists in this theory. In fact,
the finite action solutions of the 5D SU(2) theory on ${\bf R}^4\times 
S^1$
are given by two infinite towers obtained from the ordinary
instantons on ${\bf R}^4$.  The analysis of these solutions
is a generalization to 5 dimensions
of the ${\bf R}^3\times S^1$ analysis carried out in \cite{DHKM}.  There
the role of the 3D instantons was played by the BPS monopoles.

The first infinite tower of instanton configurations, labeled by
$n\in {\bf Z},$
is obtained from the ordinary ${\bf R}^4$ instanton by applying
periodic gauge transformations
\begin{equation}
U_n=\exp\left(i n {x_5\over R} \sigma^3\right)
  \ . \label{uper}
\end{equation}
As a result of these gauge transformations, $\Phi \rightarrow U^\dagger
\Phi U + U^\dagger \partial_5 U$,  the large-distance
asymptotics of the $\Phi$-component of the instanton becomes
\begin{equation}
\Phi \rightarrow \sigma^3\left({A\over 2}+i{n\over R}\right)
  \ . \label{tow1}
\end{equation}
The existence of this tower represents the fact that the
ordinary instanton can wrap the extra dimension an arbitrary number
of times. It is also related to the fact that once an extra compact
dimension is added to the ordinary 4D theory, there will be additional
gauge transformations related to the existence of the extra
dimension. A summation over the whole instanton tower generated as
above will ensure that the final result is gauge invariant under the
full 5D gauge transformations, and not only under the subgroup
generated by 4D transformations.

The second tower is obtained from the first tower by applying
an anti-periodic gauge transformation,
\begin{equation}
U_{\rm special}=\exp\left(i  {x_5\over 2R} \sigma^3\right)
  \ . \label{uspe}
\end{equation}
This ``improper'' gauge transformation is not among the usual gauge
transformations of the theory, since it obeys an antiperiodic boundary
condition instead of being periodic. However,
since all the fields of the model are in the adjoint representation of
SU(2),
this gauge transformation does not change the periodicity of
the field configurations.
Therefore the instanton solution generated
this way still obeys periodic boundary conditions, and has to
be considered as an ordinary instanton solution.
The large-distance
asymptotics of the $\Phi$-component of the second instanton tower is
\begin{equation}
\Phi \rightarrow \sigma^3\left({A\over 2}+i{n+1/2\over R}\right)
  \ . \label{tow2}
\end{equation}

In order to derive the instanton contribution to $\tau$ of the 5D theory
we simply need to sum over the contributions to
$\tau_{\rm SW}$ of all the instanton configurations in each tower.
Since the contribution of a single instanton  is given by
$\tau_1 \Lambda^4/a^4$ as in in (\ref{tauexp}),
the sum over the two instanton towers is:
\begin{eqnarray}
\label{instantonsum}
&&\tau^{\rm 1-inst}(A)=
{\tau_1 \Lambda^{4} \over 2^4} \sum_{n=-\infty}^{\infty}\left(
{1\over\left({A\over 2}+i{n\over R}\right)^4}+
{1\over \left({A\over 2}+i{n+1/2\over R}\right)^4}\right)
\\
&&=  {\tau_1 \Lambda^{4}R^4\over 2^4} {1\over 6}{\partial^2 \over 
\partial
x^2}
\sum_{n=-\infty}^{\infty}\left({1\over(x+in)^2}+{1\over(x+i(n+1/2))^2}\right)
=  {\tau_1 \Lambda^{4}R^4\over 2^4} {1\over 6}{\partial^2 \over \partial
x^2}
{4 \pi^2 \over \sinh^2(2\pi x)}
\ , \nonumber
\end{eqnarray}
where we have introduced the notation  $x=AR/2.$
Combining with the perturbative expression for $\tau$
we obtain the final result:
\begin{equation}
\tau= {i\over \pi} \log \left({\sinh^2(\pi A R)\over \pi^2
R^2\Lambda^2}\right)
+ (\Lambda \pi R)^4 \tau_1\left(
{1\over\sinh^4(\pi A R)} + {2\over 3} {1\over\sinh^2(\pi A R)}
\right)
  \ . \label{tau1i1i}
\end{equation}
Comparing this to \eqref{tau1ic} and using $\tau_1=-(i/\pi) 3G_1$
we confirm the prediction of the 5D curve
and in addition  fix the 1-instanton coefficient in the function $f$
\begin{equation}
f_1=-4G_1=-1 \ .
\end{equation}
The consistency of the exact result with the explicit instanton 
calculation
is strong evidence for the absence of
fully localized 5D instantons with finite action.
Such instantons would give additional contributions to $\tau$,
which we do not see.
Furthermore, the agreement between the curve prediction
and our instanton calculation confirms the rules
for explicit 5D instanton calculations detailed above.

\section{Non-perturbative Results from Deconstruction}
\setcounter{equation}{0}
\setcounter{footnote}{0}

In this section we will study the 5D theory using deconstruction.
A deconstructed version of a 5D theory is a 4D gauge theory. For
an appropriate choice of vevs of its fields, this 4D theory
gives a latticized version of the original 5D 
theory~\cite{deconstruction}.
The deconstructed
version of the theory discussed here was proposed in \cite{CEGK}, and
we refer the reader there for a demonstration of the perturbative
agreement of the deconstructed and continuum theories.  In what
follows we demonstrate exact nonperturbative agreement of the gauge
coupling functions in the deconstructed and continuum
theories.  The comparison between the deconstructed and the
continuum theories has to be done in the (infinitely) strong
coupling regime of the deconstructed theory. However, the
quantities that we are going to calculate are protected by
holomorphy, and thus our results remain reliable. In
addition, the deconstructed theory provides a more precise
understanding of the meaning of instanton effects in five-dimensions.

\subsection{Review of the deconstructed theory and its Seiberg-Witten
curve}

Consider the 4D ${\cal N}=1$ SU(2)$^N$ theory with bifundamental
chiral multiplets as in \cite{CEFS}.  This is the deconstructed
version of the 5D ${\cal N}=1$ SU(2) theory, as described in
\cite{CEGK}.  To be explicit, the deconstructed theory is given by
${\cal N}=1$ vector multiplets for each of the SU(2) gauge groups, and
chiral multiplets $Q_i$ transforming as summarized in the following
table:
\begin{equation}
\label{table}
\begin{array}{c|ccccc}
  &SU(2)_1&SU(2)_2&SU(2)_3&\cdots&SU(2)_N \\ \hline
Q_1&\Yfund& \overline{\Yfund}&1&\cdots&1 \\
Q_2&1&\Yfund& \overline{\Yfund}&\cdots&1 \\
   &\vdots&\vdots&\vdots&\ddots&\vdots   \\
Q_N&\overline{\Yfund}&1&1&\cdots&\Yfund
\end{array}
\end{equation}
The gauge invariant operators (whose vevs parametrize the moduli space) 
are
$B_i={\rm det}\  Q_i,$ $i=1,\dots,N$ and $T={\rm Tr}\ (Q_1\cdots Q_N).$
The Seiberg-Witten curve
for the product group theory is most easily expressed \cite{CEFS}
in terms of a composite
field which transforms as an adjoint under one of the SU(2)'s, namely,
\begin{equation}
\label{Phidef}
\Phi=Q_1Q_2\cdots Q_N-\frac{1}{2}{\rm Tr}\ (Q_1Q_2\cdots Q_N).
\end{equation}
 From this adjoint we form the usual SU(2) invariant vev, $\tilde{u}=\,
\left<{\rm Tr}\,\Phi^2\right>$,
which is then re-expressed in terms of the gauge invariants $T$ and 
$B_i$,
taking into consideration the quantum modified constraints among gauge
invariants.
The Seiberg-Witten curve is then given by \cite{CEFS},
\begin{equation}
\label{deccurve}
y^2=\left(x^2-\tilde{u}\right)^2 -4\prod_{j=1}^N\Lambda_i^4 
\end{equation}
This has the
form of the 4D ${\cal N}=2$ Seiberg-Witten curve in terms of the modulus
$\tilde{u}$.  This curve was shown to agree with a brane picture of the
theory
in \cite{LPT}.
To compare with the 5D theory we first give identical
vevs $v{\bf 1}$ to the $Q_i$,
and we assume all the couplings and $\Lambda$'s are equal.  The vevs
break the SU(2)$^N$ theory to a diagonal $SU(2)$.  The corresponding
5D theory (classically) has a
lattice spacing $l=1/gv$ and a radius $R=
N/(2 \pi g v)$, where $g$ is the gauge coupling of the individual
SU(2) factors.  This identification is most easily determined by 
comparing
the spectra of the deconstructed and continuum theories
\cite{deconstruction}.
However, the exact Seiberg-Witten results are most easily written
in terms of holomorphic quantities.  In particular, it is the holomorphic
NSVZ gauge coupling \cite{NSVZ} that is relevant here.
This requires that
the normalization of the fields be changed from the one conventionally
used in deconstructed models, and should instead coincide with
the normalization used in the preceding sections.
We can accomplish this by redefining
the gauge fields as $A_{\mu}'= g A_{\mu}$, so that the gauge kinetic
terms in the new variables become $-\frac{1}{4 g^2} 
F_{\mu\nu}F^{\mu\nu}$.
Since in the deconstructed theory in the limit $N\to \infty$ one expects
to recover ${\cal N}=2$ supersymmetry in 4D \cite{CEGK},
one  needs to rescale
the scalar fields and the fermions as well, such that, for example, the
bosonic kinetic term becomes:
\begin{equation}
\label{kinterm}
{\cal L}_{kin}=-\frac{1}{4 g^2}  F_{\mu\nu}^i F^{\mu\nu ,i}+\frac{1}{g^2}
D_\mu Q_i^\dagger D^\mu Q_i,
\end{equation}
where the covariant derivative is now given by
$D_\mu \varphi =(\partial_\mu -iA_\mu^a T^a) \varphi$.
In fact from the derivation in \cite{IS,CEFS} of the Seiberg-Witten
curve (\ref{deccurve}) it is easy to see that even for finite $N$ the
moduli
in the curve are implicitly defined in terms of the rescaled fields
with the kinetic term given by (\ref{kinterm}).

In this normalization we then obtain the holomorphic gauge coupling.
However, the usual formula for the radius of the deconstructed extra
dimension has to be modified. The reason is that in this
normalization the physical masses of the gauge bosons
are changed to $4 v^2 \sin^2 \frac{n\pi}{N}$, where $v$ is the
vev of the rescaled scalar bifundamentals. Therefore the lattice
spacing is given by $l=1/v$, and the radius of the extra
dimension is $R=N/2\pi v$. One can see that this radius is holomorphic
in the fields, as required from a quantity that we expect to appear
in the SW curve. We will refer to this radius as the holomorphic
radius.
Notice that at this point the radius is defined perturbatively.
In particular, the spectra through
which the radius is defined are expected to receive nonperturbative
corrections.
By studying the Seiberg-Witten
curve and explicit instanton contributions to the moduli of the
deconstructed
theory we will be able to make a precise nonperturbative definition of 
the
radius of the 5D theory.

\subsection{Matching of the perturbative mass spectra}
\label{pertmass}

Once we higgs the theory
down to the diagonal subgroup with a vev proportional to the identity
for each of the bifundamentals $Q_i$,
we can shift the vevs of $Q_i$
by an amount proportional to $\sigma_3$
in order to give a vev to the adjoint of the
5D theory. The shifted vevs break the gauge group to a single $U(1)$.
Furthermore,
notice that giving the same diagonal vev to all the $Q_i$ also satisfies
the
$D$-flatness constraints, \begin{equation} Q_iQ_i^\dagger-
Q_{i+1}^\dagger Q_{i+1}\propto \mathbf{1}.
\end{equation}
Hence,
we have, \begin{equation}
\label{vplusminus}
Q_i=\left(\begin{array}{ll}
v_+ & \\
& v_- \end{array} \right).
\end{equation}

Let us first match the perturbative mass spectrum of the gauge bosons of
the
deconstructed theory to that of the 5D theory. This is obtained by
analyzing the kinetic terms for the bifundamental scalars.
The covariant derivative on the bifundamental will be given by,
\begin{equation}
D_\mu Q_i = \partial_{\mu} Q_i -\frac{i}{2} \left(\begin{array}{ll}
  A^{(i)}_\mu   &  \sqrt{2} W^{(i)-}_\mu \\
\sqrt{2} W^{(i)+}_\mu & - A^{(i)}_\mu\end{array} \right) Q_i+
\frac{i}{2} Q_i \left(\begin{array}{ll}
  A^{(i+1)}_\mu   &  \sqrt{2} W^{(i+1)-}_\mu \\
\sqrt{2} W^{(i+1)+}_\mu & - A^{(i+1)}_\mu\end{array} \right),
\end{equation}
where $A^{(i)}$ denotes the third gauge boson of the $i^{th}$ gauge 
group,
while $W^{(i)\pm}=(A^{(i),1}\pm A^{(i),2})/\sqrt{2}$. Substituting the
vev of $Q_i$ into the kinetic terms we obtain a mass term for the
gauge bosons of the form,
\begin{equation}
\frac{1}{4} \sum_i \left[ \left((A^{(i+1)}-A^{(i)})^2+4 |W^{(i)}|^2
\right)(|v_+|^2+|v_-|^2)- 4  (W^{(i+1)+} W^{(i)-}
v_+v_-^* +h.c.) \right].
\end{equation}
This will give rise to a mass matrix for the $A$ bosons of the form,
\begin{equation}
\frac{ (|v_+|^2+|v_-|^2)}{2} \left( \begin{array}{rrrr}
2 & -1 & & -1 \\
-1 & 2 & \\
& & \ddots & -1 \\
-1 & & -1 & 2 \end{array} \right).
\end{equation}
The mass eigenvalues are then given by
\begin{equation}
\label{massofw3}
m_n^2= 2 (|v_+|^2+|v_-|^2) \sin^2 \frac{\pi n}{N},
\end{equation}
from which the
radius of the extra dimension in the large $N$ limit is
read off to be $R=\frac{N}{\pi \sqrt{2 (v_+^2+v_-^2)}}$, and  the
corresponding lattice spacing is given by $a^{-1}= (v_+^2+
v_-^2)/2)^{\frac{1}{2}}$.
The masses of the W bosons are given by the matrix
\begin{equation}
\frac{1}{2} \left( \begin{array}{cccc}
C & -B & & -B^* \\
-B^* & C & \\
& & \ddots & -B \\
-B & & -B^* & C \end{array} \right),
\end{equation}
with $C=2 (|v_+|^2+|v_-|^2)$ and $B= 2 v_+^*v_-$.
The mass eigenvalues of the W bosons are then given by
\begin{eqnarray}
\label{massofw}
\tilde{m}_n^2& = &|v_+|^2+|v_-|^2 -v_+^* v_- e^{\frac{2\pi in}{N}}
-v_+v_-^* e^{-\frac{2\pi in}{N}}=|v_+-v_-e^{\frac{2\pi in}{N}}|^2
\nonumber \\
&=& m_n^2 +
|v_+-v_-|^2 \cos \frac{2 \pi n}{N}+i (v_+v_-^*-v_+^*v_-)\sin
\frac{2 \pi n}{N}.
\end{eqnarray}
In the large $N$ limit this reduces to $n^2/R^2 + |v_+-v_-|^2$, which
has to match the expression in the continuum limit
in order to match the expectation value of the 5D adjoint field
correctly. The corresponding expression for the mass of the KK modes
in the continuum theory in terms of the adjoint vev $A$ is
$\tilde{m}_n^2=n^2/R^2 + A^2$.
 From this we obtain that $v_+-v_-=A$.

We should comment on the fact
that the large-N perturbative spectrum agrees with that of the 5D theory
for
fixed values of the $N$ extra moduli (one linear combination of  the
$N+1$ moduli $T,  B_1, ...,  B_N$ is the $SU(2)_D$ modulus).
There are several possible ways to deal with the extra $N$ moduli. For
example,
in the brane construction reviewed in the next Section (\ref{branepic}),
the $N-1$ anomalous $U(1)$ symmetries are gauged (anomalies are
cancelled via Green-Schwarz mechanism at the cutoff scale). Their D-flat
conditions now leave only
2 moduli, $T$ and $B_1 ... B_N$.
One combination of the two is then  the $SU(2)_D$ modulus. The real part 
of
the
remaining modulus  can be interpreted as the radion
  of the compactified continuum 5D theory, while its imaginary
part can  be identified with  the Wilson line of the graviphoton $B_5$.  
It
is possible to
stabilize the remaining modulus by adding a Lagrange multiplier term for
$B_1 ... B_N$ to the superpotential.
In the continuum theory, this term would have the interpretation as 
arising
due to some (unspecified) radion
stabilization mechanism. Alternatively, without employing anomalous
$U(1)$s,
one could  stabilize all baryons via Lagrange multipliers $L_i$, e.g. by
adding a
superpotential of the form $W = L_i ( B_i - v^2)$.

\subsection{Correspondence between continuum and deconstructed 
instantons}

\label{branepic}

We showed that the perturbative spectra of the compactified continuous
  and large-$N$ deconstructed theories agree. The next step towards
  demonstrating the equivalence of the two theories is to find a map
  between the (semiclassical) nonperturbative effects.  In this
  section, we will discuss in some detail the map between instanton
  contributions to the low-energy $\tau$ parameters in the two
  theories.

On the compactified 5D theory side,
the semiclassical calculation of the instanton
corrections to the ``photon"  $\tau$ parameter involves a sum over two
towers of instantons.
These two towers of instanton solutions are obtained from the
four-dimensional
BPST instanton by applying the ``proper" periodic (\ref{uper})  and
``improper," i.e. antiperiodic (\ref{uspe}),  large gauge 
transformations.
These transformations only exist
in the unbroken $U(1)$ subgroup of the  $SU(2)$  theory on $S^1$ since
$\pi_1 (U(1)) = Z$, while  $\pi_1(SU(2)/U(1))=  \pi_1(SU(2)) = 0$.
The summation
over these towers of instantons ensures that the instanton amplitude is
gauge invariant. In other words, the full gauge invariance of the 5D 
theory
is recovered only after all the semiclassical configurations in each
instanton tower are taken into account.

Now let us consider instanton configurations in the deconstructed theory.
This is a four-dimensional product group theory and its instanton 
solutions
are given by the complete set of instantons in each of the $SU(2)$ gauge
factors. The general instanton solution of this theory is a
$(k_1,k_2,\ldots,k_N)$-instanton, where $k_i$ stands for an instanton
charge
in the $i^{\rm th}$ $SU(2)$ gauge factor.

In order to establish the correspondence between instantons in the
two theories,
we have to identify the contributions of the two instanton towers of
charge $k$ in the continuum 5D theory,
with the contributions of the diagonal
$(k,k,\ldots,k)$-instanton in the product group theory in the large $N$
limit.
At the same time, the off-diagonal or so-called fractional instantons,
$(k_1,k_2,\ldots,k_N),$ with $k_i \neq k_j$ have no semiclassical 
analogues
in the continuum 5D theory.
The argument in favour of such an identification is as follows:

{\bf 1.} In the following section we will derive
the matching of the dynamical scales of two theories, \eqref{matchl},
which identifies an instanton charge $k$ in 5D with
$N^{-1}\sum_{i=1}^N k_i$ in 4D.

{\bf 2.} The instanton in the deconstructed theory should break the
diagonal $SU(2)_D$ subgroup in order to be compared to the instanton
in the continuum 5D theory in the Coulomb phase.
This requirement together with {\bf 1}
singles out the $(1,1,\ldots,1)$-instanton
as the counterpart of the $k=1$ instanton in 5D.

We now discuss the analogs of the large gauge transformations
(\ref{uper})  and  (\ref{uspe})
in the deconstructed theory and their relation to the instanton calculus.
An instructive way to find the   large gauge transformations is via the
brane construction of
  the four dimensional $SU(2)^N$ theory  \cite{LPT}. An added bonus of the
brane picture is the
  simple geometric interpretation of the deconstructed KK mass spectrum.

  The brane-engineered deconstructed theory is a $C^2/Z_N$
orbifold of the type-IIA  construction of pure ${\cal{N}}=2$ $SU(2 N)$
theory of ref.~\cite{HW}. It
involves $2 N$ D4-branes,  with world volumes in $x^0... x^3$ and $x^6$,
  suspended between two parallel NS5 branes  with world volumes in $x^0...
x^5$ and
separated along $x^6$. The orbifold acts on the $x^4 + i x^5$ (as well as
on $x^6 + i x^7$)
coordinates; the details are given in \cite{LPT}.

  \begin{figure}
\PSbox{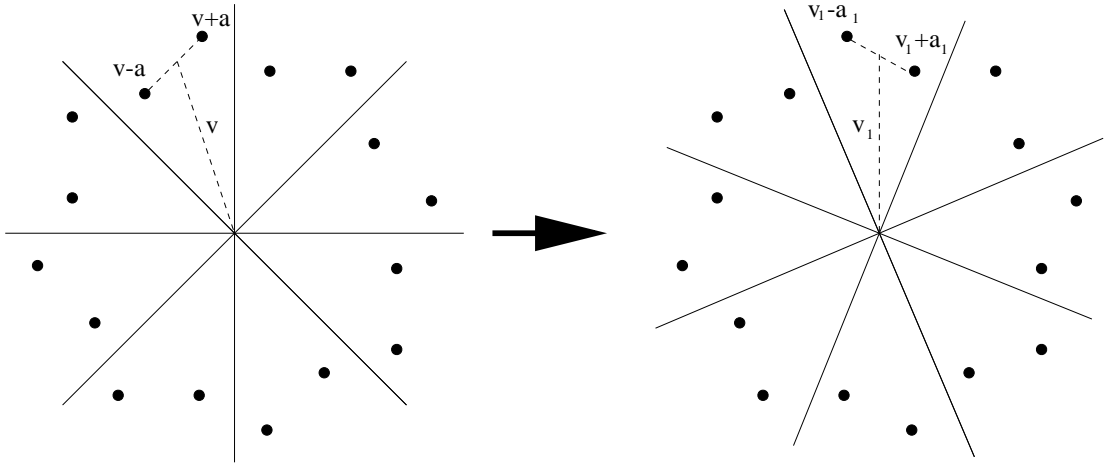 hscale=60 vscale=60 hoffset=30  voffset=0}{8cm}{5cm}
\caption{The classical moduli space of the $SU(2)^N$  theory (shown for
$N=8$) and the
$k = 1$  large gauge
transformation in the brane construction.}
\label{fig:transform}
\end{figure}

What is important for us is the description of the classical moduli
space of the orbifold theory. The $2N$ D4 branes are only allowed to move
in the
$x^4 + i x^5$ plane, in a $Z_N$ symmetric manner, as shown in Figure 1. 
The
most
general configuration is that of two branes in each $Z_N$ wedge, away 
from
each
other and from the origin. As indicated in the figure, one can identify 
the
positions
of the two branes with the parameters $v_{\pm}$ of (\ref{vplusminus}). 
The
center of
mass of the two branes in a given $Z_N$ wedge is identified with the vev
$v$,
  breaking $SU(2)^N$ to the diagonal group, while the relative 
displacement
is the
expectation value of the diagonal-$SU(2)$  adjoint field, i.e. $2 a = A$.
In particular,
the mass spectrum given in (\ref{massofw3}),  (\ref{massofw}) can be 
easily
derived from the picture.
The KK masses in the deconstructed theory are  given by  the lengths of
the strings stretched
  between the branes in a  given $Z_N$ wedge and their images. For 
example,
in the
simplest case of unbroken $SU(2)_D$, $v_{+} = v_{-} = v$,  the length 
of a
string stretched
  between a brane a distance $v$ from the origin and its $k$-th image is
  $m_k = 2 v \sin {k \pi \over N}$, as in (\ref{massofw3});
the masses in the broken $SU(2)_D$ vacuum (\ref{massofw3}, \ref{massofw})
can also
be easily derived from the geometry of the brane construction.
In a picture where all $Z_N$ wedges are identified, the
  deconstructed KK modes correspond to open strings winding around the
cone.\footnote{ It is
interesting to note  that the brane picture suggests  that  string theory
T-duality may be
underlying deconstruction, at least in the supersymmetric cases.
To see this, note that
the  large-$N$ limit of Figure 1 looks like a continuous distribution of
branes on a circle of radius  $v$
(in string units; recall that $1/v$ is the size of the UV cutoff in the
deconstructed 5D theory).
The distance between two neighboring branes is
    $\simeq 2 \pi  v/N$ (in string units and at large $N$).
T-duality relates a straight infinite periodic chain of D$p$ branes, with
period $2 \pi v/N$,  to a D$(p+1)$
brane with worldvolume wrapped on a circle of circumference $2\pi R = 
N/v$.
The worldvolume theory of the latter
is a compactified $p+1$-dimensional Yang-Mills theory (the use of 
T-duality
to the construction on Fig. 1
can  be strictly justified only  in the $v \rightarrow \infty$  limit).}

It is important to note that there is a discrete arbitrariness  in the
assignment of
pairs of branes to $Z_N$ wedges in this  picture. As we will see, one can
regroup the branes into
pairs in $N$ different $Z_N$ invariant ways, one of which is shown on
Figure 1. One can pair a brane
in a given wedge with the image of the other brane in the neighboring 
wedge
and then
redraw the $Z_N$ wedges to pass between the original pair.
The ``old" and ``new" wedges are shown on the left and right in Figure 1,
respectively.
The resulting world volume theory is, of course, identical to the 
original
one in all aspects,
including masses and interactions.

It is easy to work out the transformation corresponding to the regrouping
shown
on the Figure in terms of $v_{\pm}$:
from the  picture one can immediately see that the relation between
$v_{\pm}$  (the vevs in
the ``old" wedge) and $v_{1, \pm}$ (the vevs in the ``new" wedge) is:
\beq
\label{transform1}
v_{1, +} = \alpha^{-1}  v_{-}~, ~ ~ v_{1, -} = v_{+}~,
\eeq
where $\alpha = e^{i 2 \pi/N}$. Clearly, one can generalize this 
regrouping
in $N$ different
$Z_N$ symmetric ways, by combining one of the branes in the 1st wedge 
with
the image of the other brane in the k-th (counting clockwise) wedge. The
resulting transformation is:
\beq
\label{transform2}
v_{+} \rightarrow \alpha^{-k} v_{-}~, ~~ v_{-} \rightarrow  v_{+}~,
\eeq
with $k = 1,...,N$; the transformation with $k=N$ gives, of course, the
original pair.

It is clear from the mass formulae (\ref{massofw3}), (\ref{massofw})
  that the mass spectrum is invariant under the transformations
(\ref{transform2}): the masses of the KK tower
of the vector supermultiplet, neutral  under the diagonal $U(1)$,
  are invariant, while the the transformations with $k\ne N$ shift  the KK
number of
the $W^{\pm}$  vector supermultiplets by $k$ units. It is easy to see 
that,
in the large-N limit,
  the action of the transformation (\ref{transform2})
  on the spectrum is exactly that of the continuous large gauge
transformations (\ref{uper}, \ref{uspe}).
At large $N$ and fixed $R$, recalling $v = N/(2 \pi R)$, 
(\ref{transform2})
reduces to:
\beq
\label{transform3}
v \rightarrow v~, ~~ a \rightarrow - a -  {i k \over 2 R}~.
\eeq
The minus sign can be undone by a transformation in the Weyl group,
$a \rightarrow - a$ (or equivalently, by accompanying  (\ref{transform2})
with an
interchange of $v_{+}$ and $v_{-}$). Hence,
recalling the identification $a = A/2$, we see that the action of both 
the
proper and improper (\ref{uper}, \ref{uspe})
continuum large gauge transformations is reproduced by the deconstructed
theory, for even and odd $k$, respectively.

It is possible
to construct the discrete transformations giving rise to 
(\ref{transform2})
directly in the field theory. The ones with even $k$ correspond then to
gauge transformations, while those with odd $k$ are ``improper"
gauge transformations, in one to one correspondence with the continuum
theory. It is easy to check that
both  types of  large gauge transformation  are symmetries of the
deconstructed theory action.

Instantons can now be easily added into the brane picture of the
deconstructed
theory. In fact, an instanton of the type $(1,1,\ldots,1)$ corresponds
to a D0-brane in the vicinity of each of the $N$ pairs of D4-branes.
In other words, there is a D0-brane in each of the $N$ wedges
depicted in Figure 1.
Now, we can redraw the wedges in exactly the same way as above and
discover that there is still precisely one D0 brane inside each new 
wedge.
Of course its position inside  the wedge has changed,
but we need to integrate
over the D0-brane positions when we calculate instanton partition
functions.
Integrations over instanton collective coordinates (bosonic and 
fermionic)
in field theory correspond to integrations over the D0-brane
positions in each wedge.
This means that the integral over the  $(1,1,\ldots,1)$-instanton measure
is automatically invariant under  \eqref{transform2}.
This transformation is a symmetry not only of the microscopic action,
but also of the D-instanton theory. In the deconstructed theory
there is no need to sum over the instanton images under 
\eqref{transform2}.

We can understand the difference between the continuum and the
deconstructed case in more detail by considering what the gauge
transformations in these two theories are. In the continuum calculation 
we
have viewed the theory from the effective 4D theory's point of view.
This means that all information about the 5$^{th}$ coordinate
in that theory was lost, all we kept was a tower of 4D KK modes. Then we
have
considered the 1-instanton in this effective theory. Since we omitted
the $x_5$ dependent gauge transformations from the effective theory, the
instanton measure and action will not be invariant under the large gauge
transformations. In order to reproduce the correct 5D answer, this
additional
symmetry has to be imposed by hand, which is achieved by the summation
over the two towers of the gauge-transformed instantons. The analog of
this procedure in the deconstructed theory would be to take
the 1-instanton in
the unbroken diagonal $SU(2)_D$ gauge group. This instanton (and its
measure)
would not be invariant under all the broken $SU(2)$ gauge groups, and a
way to restore the full gauge invariance would be to sum over the
discretized versions of the large gauge transformations described above.
However, a more natural way to proceed in the deconstructed theory
is to consider the effect of the $(1,1,\ldots ,1)$ instanton. In this 
case,
the situation is very different from before. The main difference is   
that,
as explained above, the discretized version of the $x_5$ dependent
gauge transformations are themselves part of the gauge symmetries of the
theory, they are simply given by $i$ dependent gauge transformations
in the $SU(2)_i$ factors. Also, as explained above, instead of 
considering
a single instanton, one would have to look at the $(1,1,\ldots ,1)$
instanton
calculation, and thus in effect calculate an $N$ instanton amplitude.
However, the  $N$ instanton measure must be constructed in  a way
that it is completely gauge invariant. Thus, there would be
no need for additional summation over the images of the
$(1,1,\ldots ,1)$ instanton, that sum is implicitly performed
by using the correct $N$-instanton measure for the theory.
Hence we conclude that the contribution of the  
$(1,1,\ldots,1)$-instanton
in the large $N$-limit must match the contribution of the
two 1-instanton towers in the continuum theory.

This argument applies directly to all diagonal
$(1,1,\ldots,1)$-instanton effects.
We have thus constructed a dictionary relating the $SU(2)_D$
instantons, contributing to the $\tau$ parameter in the deconstructed
theory
to those in the continuum theory.

\subsection{Deriving the continuum Seiberg-Witten curve from
deconstruction}

Given the identification of the instantons in the continuum and
deconstructed theories, we are now ready to compare the Seiberg-Witten
curves for the two theories. We should stress again that the
deconstructed theory has only four supercharges, while the continuum
theory has eight.
Therefore, a priori, the curve obtained through deconstruction
contains less information than the original Seiberg-Witten curve
(or Nekrasov's curve).
With eight supercharges one can exactly solve both
for the K\"ahler potential and the gauge kinetic function, while in
this case only the gauge kinetic function can be obtained.\footnote{See,
however, the discussion at
the end of Section \ref{specialflat}.}

As explained before, in order to obtain
the Seiberg-Witten curve for the deconstructed
theory one needs to evaluate $\tilde{u}=\langle {\rm Tr} \Phi^2
\rangle$, with $\Phi$ given in (\ref{Phidef}).
Using $Q_i = {\rm diag} (v_+,v_-)=(v+A/2,v-A/2)$
we can now write $\Phi$ classically as:
\begin{eqnarray}
\Phi&=&\left[
v_+^N-v_-^N\right]\,\frac{\sigma_3}{2}
=v^N\left[\left( 1+\frac{A}{2 v} \right)^N
-\left( 1-\frac{A}{2 v}\right)^N\right]\,\frac{\sigma_3}{2} \\
&=&
v^N\left[\left(1+\frac{\pi R A }{N}\right)^N-
\left(1-\frac{\pi R A }{N}
\right)^N\right]\,\frac{\sigma_3}{2} \\
&\rightarrow&v^N\,\sinh(\pi R A )\,\sigma_3.
\end{eqnarray}
Here we have used the holomorphic radius $R=N/2\pi v$.
This corresponds to the radius that appears in Nekrasov's curve
(\ref{curve5d}), since this is the correct holomorphic variable.
We also have, \begin{equation}
\tilde{u}=\left< {\rm Tr}\,\Phi^2\right>
\rightarrow \left<2 v^{2N}\,\sinh^2(\pi R A)\right>.
\label{eq:utilde}
\end{equation}
Thus we can see that $\tilde{u}$ includes
the correct variable of the 5D curve in the continuum limit. The
appearance of the gauge invariant
$\sinh^2(\pi R A)$  in the 5D curve is predicted
from the deconstructed theory.

In order to actually
match the deconstructed curve to the 5D curve obtained
above, we have to first calculate the relation between the scale 
$\Lambda$
appearing in the deconstructed curve (\ref{deccurve}) and the low-energy
scale $\Lambda_D$ which appears in the 5D curve. The matching is
slightly non-trivial due to the presence
of the KK modes, whose effects on the running of the
coupling have to be taken into account. The matching of the
holomorphic gauge
couplings at the scale of the highest KK mode $m_{KK}=2 v$ is
given by
\begin{equation}
\frac{1}{g_D^2}=\frac{N}{g^2}\ .
\end{equation}
We now want to run the diagonal coupling down to a scale $\mu$ which is
below the mass of the lowest KK mode. The renormalization group
evolution equation is
given by
\begin{equation}
\frac{1}{\alpha_D (\mu )} = \frac{N}{\alpha (m_{KK})} - \frac{2}{\pi} 
\log
\frac{m_{KK}}{\mu} -\frac{2}{\pi} \sum_{n=1}^N \log \frac{m_{KK}}{m_n}\ ,
\end{equation}
where the first logarithm is the effect of the zero modes, while the
sum gives the contribution of the KK modes, and $\alpha = g^2/4\pi$.
The mass ratio
in the logarithm is just given by $\frac{m_{KK}}{m_n} =\frac{1}{\sin
\frac{n \pi}{N}}$. Using the relation \cite{CKT}
\begin{equation}
\prod_{n=1}^{N-1} \sin^2 \frac{n\pi}{N} = \frac{4N^2}{2^{2N}}  ~,
\end{equation}
we obtain the expression for the low-energy gauge coupling
\begin{equation}
\frac{1}{\alpha_D (\mu )} = \frac{N}{\alpha (m_{KK})} - \frac{2}{\pi} 
\log
\frac{m_{KK}}{\mu} +\frac{1}{\pi} \log \frac{4N^2}{2^{2N}}\ .
\end{equation}
Using the definitions of the scales
\begin{equation}
\label{scales}
\Lambda^4_D=\mu^4 e^{-\frac{8\pi^2}{g_D^2 (\mu )}}\ , \ \ \
\Lambda^4=m_{KK}^4 e^{-\frac{8\pi^2}{g^2 (m_{KK})}}\ ,
\end{equation}
we obtain the scale matching relation
\begin{equation}
\Lambda^4_D=\frac{\Lambda^{4N}}{m_{KK}^{4N-4}} \frac{2^{4N}}{16 N^4}\ .
\end{equation}
Using $m_{KK}= 2 v$ and $2\pi R=N/v$ this can be rewritten as
\begin{equation}
\label{matchl}
  \Lambda^4_D=\frac{\Lambda^{4N}}{v^{4N}} \frac{1}{(2 \pi R)^4}\ .
\end{equation}
There may a priori be instanton corrections to these matching
relations, but we can make precise the correspondence between the
parameters
of the deconstructed and continuum Seiberg-Witten curves as follows.

First, we define a $Z_N$ symmetric
gauge invariant radius (along the branch of moduli space
where this identification makes sense) via  \begin{equation}
  \left({N \over
2\pi R} \right)^{2 N} \equiv
\prod_{i=1}^N B_i \ .
\end{equation}
In the continuum limit along the branch of moduli space we are 
considering
$B_i\rightarrow v^2$.\footnote{Recall that in this limit,
$A/v \rightarrow 0$, so that $v_\pm\rightarrow v$.}  
For simplicity we define $B=(\prod_i B_i)^{1/N}$.
Let us now rescale the curve in (\ref{deccurve}) by
$x^2\to x^2 B^{N} (2\pi R)^2$ and $y^2\to y^2  B^{2N}
((2\pi R)^2)^2$,
and rescale the modulus by, \begin{equation} \label{eq:ubar}
\overline{U} \equiv
{\tilde{u} \over B^{N} (2\pi R)^2 } \rightarrow {\left<2 \sinh^2 (\pi R
A)\right>\over
(2\pi R)^2}~^À  .
\end{equation}
The last relation in (\ref{eq:ubar}) deserves some comment.  It is 
obtained
by identifying $\left<v^{2N}\right>\sim \left<v^2\right>^N$.
We will demonstrate in the next section that there are no corrections to
(\ref{eq:ubar}) from instantons in the broken gauge groups.  There may be
diagonal instanton corrections to this relation (which we do not
calculate),
which may be related to
the function $f(\pi R \Lambda)$ in (\ref{funf5d}).
In what follows
the first relation in (\ref{eq:ubar}) should serve as the definition of
$\overline{U}$, which is then unambiguous.
In the large $N$ limit we can also rewrite \ref{matchl} in using
the gauge invariant definition of the radius
\begin{equation}
\Lambda_D^4 = \frac{\Lambda^{4N}}{B^{2N} \,(2\pi R)^{4}}\, .
\end{equation}
The curve we obtain then is given by
\begin{equation}
\label{5Dlimit}
y^2= (x^2-\overline{U})^2-4 \Lambda_D^4\ ,
\end{equation}
Finally, we note that in the continuum limit
$\overline{U}$ is related via (\ref{eq:ubar})
to the modulus that appears
in the continuum curve (\ref{curve5d}), and
$\Lambda_D$ is the dynamical scale in that theory;
hence, we exactly
reproduce the expected gauge coupling $\tau(\overline{U})$ in the 
continuum
theory.  In fact, to be more precise the modulus that appears in the
continuum theory in \cite{Nekrasov}
involves $\left<\cosh(\pi R A)\right>^2-1$ and in the deconstructed 
theory
it
is $\left<v^{2N}\sinh(\pi A R)^2\right>/\left<v^2\right>^N$.
Hence, deconstruction leads us to
suspect that the origin of the function $f(\pi R\Lambda)$ in 
(\ref{funf5d})
are
the diagonal instantons that relate these moduli.
Note that this function can not
be fixed by symmetry arguments,
but an explicit
instanton calculation of the sort we have performed is necessary
to determine it at every instanton level. However, this
possibility implies that matching of additional operators
between deconstructed and continuous theories may be rather
non-trivial. In the section \ref{specialflat} we will argue that
the correspondence between deconstructed and continuum
models may be more direct along certain special flat
directions of the deconstructed theory.

\subsection{The role of instantons in the broken groups and of
the quantum modified constraints}

In the following we clarify one subtlety:
the role of quantum modified constraints
in the relation between moduli of the deconstructed and continuum 
theories.

The modulus $\overline{U}_{cl}$ defined in terms of the moduli $T$ and
$B_i$ via classical constraints, and the modulus $\overline{U}$ that
becomes the modulus of the continuum theory in the appropriate limit,
differ by instanton contributions even though they have the same
classical limit. So the question is which modulus to equate with the
continuum modulus in the continuum limit.
We first answer this by a
physical argument, and then demonstrate that it is correct by a technical
one.

\subsubsection{Relations between moduli}
The continuum variable $U$ in (\ref{eq:U}) was defined in the low-energy
effective 4D theory, where the only instantons that exist
are the usual 4D $SU(2)$ instantons. However, in the deconstructed theory
there is more than just one kind of instanton. Before breaking the 
diagonal
$SU(2)$ group to $U(1)$ there are two types of instantons: the instantons
in the diagonal unbroken $SU(2)$, which will be mapped to the
instantons that remain in the effective 4D theory, but there are also
instantons in the broken $SU(2)$ factors. We can denote these as
$(1,0,\ldots ,0)$,  $(0,1,0, \ldots ,0)$ instantons, while the
instanton in the diagonal $SU(2)$ factor is the $(1,1,\ldots ,1)$
instanton \cite{CM}. Since the instantons in the broken gauge groups
have no analogs in the effective 4D theory, the definitions of the
two variables $\overline{U}$ and $\overline{U}_{cl}$ may differ by the
effects of these instantons.  To highlight the issue, we write the
deconstructed curve in terms of the moduli $T$ and $B_i$ as in 
\cite{CEFS} (along
the flat direction (\ref{vplusminus})):
\begin{equation}\label{eq:fullcurve}
y^2=\left[x^2-\overline{U}_{cl}(T,B_i)+\sum_{j=1}^N { \Lambda_j^4 \over
(2\pi R)^{  2} ~B^2} \right]^2 -4 \Lambda_D^4.
\end{equation}
So it is important that it is $\overline{U}$ and not
$\overline{U}_{cl}$ that corresponds to the modulus in the continuum 
curve
(\ref{curve5d}).
We can understand why this is the case as follows.

For the purpose of demonstration we study
the simple case of $N=2$, with the discussion easily
extended to higher $N$. For $N=2$, the theory is given by
\begin{equation}
\label{special}
\begin{array}{c|cc|c}
& SU(2) & SU(2) & SU(2) \\ \hline
Q_{aAf} & \Yfund & \Yfund & \Yfund \\
\end{array},
\end{equation}
where one has an additional $SU(2)$ global symmetry in the special case
$N=2$,
which is the last $SU(2)$ factor in (\ref{special}). This is the theory
considered by Intriligator and Seiberg in \cite{IS}, and the
derivation of the relation between moduli for this case is basically
already
contained in \cite{IS}. Here we repeat it in order to make
the argument complete, and also to give a more physical explanation
for the origin of these extra terms in (\ref{eq:fullcurve}).
The argument (which in fact is the
essence of the whole derivation of the curves in \cite{IS} and 
\cite{CEFS})
is as follows. Consider the case when the first gauge group is much
stronger than the second one, $\Lambda_1 \gg \Lambda_2$. Then the second
gauge group can be neglected and the first gauge group is simply
an $SU(2)$ theory with two flavors (four fundamentals). This theory
was described in \cite{Seiberg1} (see also \cite{otherqmc}).  At low
energies it is described by the confined mesons
\begin{equation}
M_{AfBg}=Q_{aAf} Q_{bBg} \epsilon^{ab}.
\end{equation}
This meson contains three singlets and an adjoint {\bf 3} under the
weakly gauged second gauge group. This adjoint is formed by the field
\begin{equation}
\Phi_A^B = \frac{1}{2\Lambda_1} M_{AfCg} \epsilon^{fg} \epsilon^{CB}.
\end{equation}
In terms of this adjoint field the theory is simply described by an
ordinary ${\cal N}=2$ $SU(2)$ Seiberg-Witten curve
\begin{equation}
\label{SWcurve}
y^2=(x^2+\tilde{u})^2-4 \Lambda_2^4.
\end{equation}
Here $\tilde{u}$ is the invariant formed from the composite adjoint field
$\Phi_A^B$
\begin{equation}
\tilde{u}=
\frac{1}{2} {\rm Tr} \Phi^2 = \frac{1}{8\Lambda_1^2} M_{AfCg} M_{BhDi}
\epsilon^{fg} \epsilon^{CB} \epsilon^{hi} \epsilon^{DA}.
\end{equation}
Notice that this agrees with our earlier definition of $\tilde{u}$ for 
the
general SU(2)$^N$ theory up to a dimensionful constant.
However, we would like to express the curve in terms of the natural
variable $u'$, which is defined as the invariant
\begin{equation}
u'= \det \tilde{M},
\end{equation}
where $\tilde{M}_{fg}= \frac{1}{2}
Q_{aAf} Q_{bBg} \epsilon^{ab} \epsilon^{AB}$.
We can now express the variable $u'$ in terms of $\tilde{u}$.
An explicit calculation shows that the relation between the two 
invariants
is given by
\begin{equation}
\Lambda_1^2 \tilde{u} + u' = {\rm Pf} M,
\end{equation}
where the Pfaffian ${\rm Pf} M$ is most easily expressed in terms of the
$SU(4)$ symmetric meson matrix (obtained by ignoring the gauge 
interactions
of the second gauge group since $\Lambda_2 \ll \Lambda_1$).  One
can translate between the two sets of indices of $M_{AfBg}$ and the
$SU(4)$ notation $M_{\alpha\beta}$ by the assignment
$(11)\to 1, (12)\to 2, (21)\to 3, (22)\to 4$. With this translation
${\rm Pf} M=\frac{1}{8} \epsilon^{\alpha\beta\gamma\delta} 
M_{\alpha\beta}
M_{\gamma\delta}$. However, the ${\rm Pf} M$ is exactly the
quantity which classically vanishes (once expressed in terms of the
underlying quark fields), but receives a one-instanton correction
quantum mechanically and yields the quantum modified constraint
\begin{equation}
{\rm Pf} M = \Lambda_1^4.
\end{equation}
The coefficient of the one-instanton contribution was fixed by Seiberg
\cite{Seiberg1}
by matching to the ADS superpotential \cite{ADS} after integrating out 
one
flavor, and by Finnell and Pouliot \cite{FP}
by a direct instanton calculation.
Using this relation we obtain
\begin{equation}
\Lambda_1^2 \tilde{u}+u'=\Lambda_1^4.
\end{equation}
The curve (\ref{SWcurve}) is now rewritten (after rescaling $x$ and $y$) 
as

\begin{equation}
y^2=(x^2-(u'-\Lambda_1^4))^2-4 \Lambda_1^4 \Lambda_2^4.
\end{equation}
This explains the extra shift in the curve due to instantons in
the first gauge group, and there is a similar shift due to
instantons in the second gauge group, and the final curve becomes
\begin{equation}
y^2=(x^2-(u'-\Lambda_1^4-\Lambda_2^4))^2-4 \Lambda_1^4 \Lambda_2^4.
\end{equation}
This derivation of the $SU(2)\times SU(2)$ curve
teaches us that the variable $\tilde{u}$ obtains a
correction from its classical value in terms of the fundamental moduli
$M_{ij}$ due to the instantons in the
individual gauge groups. These are the instantons which
after the breaking to the diagonal
gauge group become instantons in the broken gauge group. The extra
instanton terms in (\ref{eq:fullcurve})
arise due to the fact that the curve has been
expressed in terms of a variable which obtains a correction
from these instantons. We have used this expression for the curve since
these are the variables that are natural for the deconstructed theory.
However, in the continuum limit it is more convenient to work with
the variable $\tilde{u}$, in terms of which instantons in the broken
group never appear.  This modulus is directly related to the
modulus of the continuum theory because the low energy
continuum theory simply doesn't have such instantons in it.

To stress the point, the deconstructed analog of the continuum modulus
proportional to
$\left<\sinh(\pi R A)^2 \right>$ is related to $\tilde{u} \propto
\left<v^N[(1+ \pi R A/N)^N - (1-\pi R A/N)^N]\right>$ .  This gauge
invariant
vev, not being directly related to the fundamental gauge invariants
$M_{ij}$ (or $T$ and
$B_i$ in the general case), is subject
to quantum modified constraints among the moduli. When expressed in 
terms of
the ``fundamental'' gauge invariants $T$ and $B_i$
there appears to be a superfluous
term in the Seiberg-Witten curve, but this is only because of the choice
of gauge invariants in terms of which we expressed the curve, and is
not relevant for comparison with the 5D theory.  It remains to be proven
that $\tilde{u}(A)$ does not receive broken instanton corrections, and we
will demonstrate this (at least for one-instanton corrections) in the 
next
section.

\subsubsection{Explicit instanton calculation of $\tilde{u}(A)$}

In the following, we perform an explicit one-instanton calculation to
confirm
that the modulus $\tilde{u}$
does not receive any contribution from instantons in the broken groups.
{\em A priori}, a zero mode counting would allow such a term, but an 
exact
cancellation demonstrates that such terms are absent at the
$(1,0,\ldots,0)$-instanton
level.  This verifies the identification of $\tilde{u}(A)$ as the modulus
of the continuum theory.

Let us consider a  single instanton in the second
$SU(2)$ factor of the deconstructed theory \eqref{table}--the
$(0,1,0,\ldots,0)$-instanton\footnote{The contributions to $\tilde{u}$
of an instanton in the $n$-th $SU(2)$ factor does not depend on the value
of $n$ since $\tilde{u}$ involves a trace  over all bi-fundamentals.}.
The field components of this instanton
are the $SU(2)_2$ gauge field and gauginos, and the (anti)-fundamental
flavors $Q_1$ and $Q_2$ comprising fermions and scalars.
  Instanton components of all other fields are trivial.
Thus, from the perspective of the $(0,1,0,\ldots,0)$-instanton, the
product group theory \eqref{table} is equivalent to the ordinary $SU(2)$
supersymmetric QCD with $N_{\rm f}=2$ real flavors:
$Q_{1f}$ with $f=1,2,$
play the role of the anti-fundamental chiral flavors $\tilde{Q}_f$,
and $Q_{2f}$ are the fundamental chiral flavors $Q_f.$

We can now apply the standard rules of instanton calculus
to the case at hand. For calculating instanton contributions to 
$\tilde{u}$
we need three ingredients: the instanton action, the instanton components
of the (anti)-fundamental scalars, and the instanton measure.

Using conventions of \cite{MO2,MO1}, the instanton action is given by
\begin{equation}
\label{sinst}
S=\,{8\pi^2\over g^2} +2\pi^2\rho^2(|v_+|^2+|v_-|^2)
-{i\over \sqrt{2}} \left(\begin{array}{llll}
\bar{v}_+ &  \\
& \bar{v}_-
\end{array}\right)_{f}^{\,\,\dot\beta}
\mu_{\dot\beta}\, ({\cal K}_f+\tilde{\cal K}_f) \ ,
\end{equation}
where $\rho$ is the instanton size,
$\mu_{\dot\beta}=\{\mu_{1},\mu_{2}\}$  are the Grassmann collective
coordinates
of superconformal fermion zero modes and
${\cal K}_f$ and $\tilde{\cal K}_f$ are the Grassmann collective
coordinates
of fundamental and anti-fundamental fermion zero modes.
The (anti)-fundamental scalar components of the instanton read \cite{MO2}
\begin{eqnarray}
\label{qqtilde}
q^{\dot\beta}_{\ f}&=& \sqrt{x^2\over x^2+\rho^2}
\left(\begin{array}{llll}
\bar{v}_+ &  \\
& \bar{v}_- \end{array}\right)^{\dot\beta}_{\,\, f}\nonumber\\
&+&
{i\over 2\sqrt{2}}
{|x|\over (x^2+\rho^2)^{3/2}}\mu^{\dot\beta}{\cal K}_f
-{i\over 2\sqrt{2}}{\rho\over |x|}{1 \over (x^2+\rho^2)^{3/2}}
\bar{x}^{\dot\beta \beta}M_\beta {\cal K}_f \ ,
\end{eqnarray}

\begin{eqnarray}
\label{qtildeq}
\tilde{q}_{f\dot\beta}&=& \sqrt{x^2\over x^2+\rho^2}
\left(\begin{array}{llll}
\bar{v}_+ &  \\
& \bar{v}_- \end{array}\right)_{f\dot\beta}\nonumber\\
&-&
{i\over 2\sqrt{2}}
{|x|\over (x^2+\rho^2)^{3/2}}\tilde{\cal K}_f \mu_{\dot\beta}
-{i\over 2\sqrt{2}}{\rho\over |x|}{1 \over (x^2+\rho^2)^{3/2}}
\tilde{\cal K}_f M^\beta {x}_{\beta\dot\beta } \ .
\end{eqnarray}
Here  $M^\beta=\{M^1,M^2\}$ denote supersymmetric fermion zero modes,
and the Weyl indices $\dot\beta$ and $\beta$ are raised and lowered with
the
$\varepsilon$-symbols. The fermion-bilinear terms in the scalar 
components
above
arise from the Yukawa sources in the corresponding Euler-Lagrange
equations.

Finally, the instanton measure of the $SU(2)$ $N=1$ supersymmetric QCD
with $N_{\rm f}=2$ flavors is given by (cf \cite{MO2})
\begin{equation}
\label{imeasure}
\int d\mu_{\rm inst} =\,
{2^9\over \pi^2}{\mu_{\rm PV}^4\over g^4}\int d^4x_0\, \rho^3 d\rho\,
d^2 M \,d^2 \mu \,d{\cal K}_1 d\tilde{\cal K}_1 d{\cal K}_2 d\tilde{\cal
K}_2\,
\exp[-S] \ ,
\end{equation}
where $x_0$ is the instanton position and
$\mu_{\rm PV}$ is the Pauli-Villars renormalization scale,
\begin{equation}
\mu_{\rm PV}^4 \exp\left( -{8\pi^2 \over g^2(\mu_{\rm PV})} \right) =
\Lambda_{\rm PV}^4 \ .
\end{equation}
The instanton contribution to $\tilde{u}$ is given by
\begin{equation}
\label{tuic}
\tilde{u}=
\left<{\rm Tr}\,\Phi^2\right>=
  \int d\mu_{\rm inst}{\rm Tr}\,\Phi^2 \ ,
\end{equation}
where the instanton component of $\Phi$ can be found from \eqref{Phidef}
and \eqref{qqtilde}-\eqref{qtildeq}.

To simplify things a little we will now take the large $N$ limit
and hence set $v_+=v_-\equiv v.$ Then the expression for $\Phi$
takes form
\begin{equation}
\label{phic}
\Phi_{fh}=v^{N-2}\left(\tilde{q}_{f\dot\beta}q^{\dot\beta}_{\,h}
-{1\over 2} \delta_{fh} {\rm Tr}(\tilde{q}q)\right) \ ,
\end{equation}
and $\tilde{u}$ is
\begin{equation}
\label{tuic2}
\tilde{u}=  v^{2N-4} \left\langle {\rm Tr}(\tilde{q}q\tilde{q}q)-
{1\over 2}{\rm Tr}(\tilde{q}q){\rm Tr}(\tilde{q}q)\right\rangle \ .
\end{equation}

The instanton solution for $\tilde{q}q$ can be schematically written as
\begin{equation}
\label{qtqs}
\tilde{q}q = v^2 +v\mu{\cal K} +\tilde{\cal K}\mu v+vM{\cal K}+
\tilde{\cal K}Mv+ \tilde{\cal K}\mu^2{\cal K}+ \tilde{\cal K}M^2{\cal K}
+\tilde{\cal K}\mu M {\cal K} \ .
\end{equation}
Here we made explicit only the Grassmann collective coordinates and the
vevs. Notice that the first term on the right hand side of \eqref{qtqs}
is proportional to the unit matrix, $v^2\propto {\bf 1},$ and can be
dropped
as it does not contribute to either
$\Phi$ (which is traceless) or $\tilde{u}$.

Accordingly, the contributions to $\tilde{u}$ take form (cf 
\eqref{tuic2})
\begin{eqnarray}
\label{tuic3}
\tilde{u} = &\int d^2 M d^2 \mu d{\cal K}_1 d\tilde{\cal K}_1 d{\cal K}_2
d\tilde{\cal K}_2\, [
(v\mu{\cal K})(\tilde{\cal K}M^2{\cal K})e^{v\mu\tilde{\cal K}}
+(\tilde{\cal K}\mu v)(\tilde{\cal K}M^2{\cal K})e^{v\mu{\cal K}}
\nonumber\\
&+(v\mu{\cal K})(\tilde{\cal K}\mu v)e^{v\mu{\cal K}+v\mu\tilde{\cal K}}
+(\tilde{\cal K}\mu^2{\cal K})(\tilde{\cal K}M^2{\cal K})
+ (\tilde{\cal K}\mu M {\cal K})(\tilde{\cal K}\mu M {\cal K})] \ .
\end{eqnarray}
Performing the integrations over Grassmanian collective coordinates
and keeping careful track
of the raised and lowered indices of the supersymmetric and 
superconformal
zero
modes\footnote{Note that $\int d^2\mu \mu_{\dot\alpha}\mu^{\dot\beta}
=\delta_{\dot\alpha}^{\,\dot\beta}/2,$
$\int d^2\mu \mu_{\dot\alpha}\mu_{\dot\beta}
=-\varepsilon_{\dot\alpha\dot\beta}/2,$ etc.
} in \eqref{qqtilde}-\eqref{tuic3}
one discovers that the first term on the right hand side of
\eqref{tuic3} cancels against the second term, the third term is
vanishing and the fourth term cancels against the fifth term.
Thus we conclude that the total contribution of single instantons
of the $(1,0,\ldots,0)$-type to the modulus
$\tilde{u}$ vanishes. This fact is in agreement with our
identification of $\tilde{u}$ with the modulus of the continuum theory
which can receive instanton corrections only of the type
$(k,k,\ldots,k).$

We conclude this discussion with an observation that
such cancellation of the instanton contributions is specific to
$\tilde{u}.$ A modulus defined in a different way would not enjoy these
cancellations. To illustrate this point one can consider a slightly
different quantity
\begin{equation}
\label{anotheru}
\left\langle {\rm Tr}(\tilde{q}q\tilde{q}q)-
{\rm det}\tilde{q}\,{\rm det}q\right\rangle \ .
\end{equation}
Classically this is equal to
$\langle {\rm Tr}(\tilde{q}q\tilde{q}q)-
1/2{\rm Tr}(\tilde{q}q){\rm Tr}(\tilde{q}q)\rangle\propto \tilde{u},$
but there are quantum (1-instanton) corrections.
In fact, it is well-known \cite{IS} that there is a quantum-modified
constraint
in the $N=N_{\rm f}$ supersymmetric QCD, ${\rm det
M}-\tilde{B}B=\Lambda^{2N}.$
For our case of $N=2=N_{\rm f},$ the meson determinant is
${\rm det M}= {\rm det \tilde{q}}\,{\rm det q},$ and the baryons are
$\tilde{B}=\tilde{q}_1\tilde{q}_2$ and ${B}={q}_1{q}_2,$ where
$1,2$ denote flavor indices and the color indices are summed over.
The quantum-modified constraint is
\begin{equation}
\label{qmconstr}
\langle
{\rm det}\tilde{q}\,{\rm det}q\rangle =\langle \tilde{q}_1q_1
\tilde{q}_2q_2
\rangle + \Lambda^4 \ ,
\end{equation}
and \eqref{anotheru} can be written as
\begin{equation}
\label{aanotheru}
\left\langle {\rm Tr}(\tilde{q}q\tilde{q}q)-
{\rm det}\tilde{q}\,{\rm det}q\right\rangle =
\langle {\rm Tr}(\tilde{q}q\tilde{q}q)\rangle-
\langle \tilde{q}_1q_1 \tilde{q}_2q_2\rangle+ \Lambda^4 \ .
\end{equation}
Repeating the same $(0,1,0,\ldots,0)$-instanton calculation as above one
concludes that the first and the second term on the right hand side of
\eqref{aanotheru} cancel each other. But the last term,
$\Lambda^4,$ remains,  giving a non-vanishing single-instanton
contribution to \eqref{aanotheru}.

\subsubsection{A special flat direction }
\label{specialflat}

In this Section, we show the
existence of a  flat direction for which the partially broken instantons 
do not
contribute to the curve, even when
the curve is expressed in terms of the modulus $\overline{U}_{cl}(T,B_i)$
of eqn.~(\ref{eq:fullcurve}) (which,
along a generic direction, does receive corrections from the instantons 
in
the partially broken gauge groups).

This flat direction is easiest to infer from Figure 1. Recall that
in the brane picture, the positions of the
center of mass of the branes in the $k$-th $Z_N$ wedge correspond  to the
expectation values   $\langle Q_k \rangle = v_k \sigma^0$,
where:\footnote{The relation (\ref{qvevs}) holds
more generally, i.e.  the vevs of the $SU(2)_D$-breaking adjoint also 
obey
$a_k = \alpha^k a$, as is evident from the
brane picture.}
\beq
\label{qvevs}
v_k ~=~ \alpha^k ~v~.
\eeq
The $D$-flat conditions and  mass matrices
are invariant under the replacement of the expectation values
(\ref{vplusminus}) with (\ref{qvevs}). There are a few points to make 
about
the relevance of this phase choice,
  which might appear  arbitrary in the deconstructed $SU(2)^N$ field 
theory,
but is a consequence of
  the $Z_N$ symmetry of the brane configuration (it can, of course, also 
be
imposed on the field theory).
The most important point is that, along the flat direction
(\ref{qvevs}), the baryon expectation values obey $B_k = \alpha^{2 k} 
v^2$.
Recall now that the term in the curve
due to the instantons in the partially broken $SU(2)$ groups has the form
(see ref.~\cite{CEFS}):
\beq
\label{brokeninstantonscurve}
\sum\limits_{k=1}^N \Lambda_k^4~ B_1 \ldots  \hat{B}_{k-1}  \hat{B}_k
\ldots B_N ~,
\eeq
where  hats indicate that the corresponding fields are omitted and $0
\simeq N$. Let us, for the moment, assume that
  all the $\Lambda_k$ are equal complex numbers. Then
eqn.~(\ref{brokeninstantonscurve}) vanishes identically:
\beq
\label{brokencurve2}
\left ( \sum\limits_{k = 1}^N \alpha^{- 4 k} \right)  \alpha^2 
~\Lambda^4 v^{2 N  - 4}  ~=  ~0~.
\eeq
Hence, in the vacuum  (\ref{qvevs}),  the instantons in the partially
broken $SU(2)$ groups do not contribute to the curve and  $\tau$ 
parameter
of the low-energy $U(1)$.

Now we need to justify our assumption of equal phases of the 
$\Lambda_k^4$
factors (the assumption of equal couplings
is inherent to the idea of deconstruction).
To this end, note that the
$SU(2)^N$ field theory has $N$ anomalous global $U(1)$ symmetries with
parameters $\omega_k$, acting as follows:
\begin{eqnarray}
\label{globalsymmetries}
Q_k &\rightarrow& e^{i \omega_k} ~Q_k~, \\
\Lambda_k^4 &\rightarrow& e^{2 i \omega_k  + 2 i \omega_{k - 1}}
~\Lambda_k^4~, \nonumber
\end{eqnarray}
where $k = 1, \ldots, N$ and $k = 0$ is identified with $k = N$.
The transformations of $\Lambda_k^4$ reflect  the  $U(1)$ anomalies.
 From the last line in (\ref{globalsymmetries}) it follows that the 
$\theta$
parameters transform as follows:
\beq
\label{thetatransform}
\theta_k \rightarrow \theta_k + 2 \omega_k + 2 \omega_{k - 1} ~.
\eeq
It is easy to see, by writing (\ref{thetatransform}) as an $N\times N$
matrix equation, that for odd $N$ all $\theta$ parameters can
be put to zero by field redefinitions. Thus, the $\Lambda_k$ can be 
assumed
real from the very beginning, justifying our assumption
of equal phases.
In the case of even $N$, the rank of the matrix in (\ref{thetatransform})
is $N-1$ and
  there is one physical $\theta$ parameter---the
combination:
\beq
\label{thetaphysical}
\theta_{phys} ={1\over N}~ \sum\limits_{k = 1}^{N} (-1)^{k + 1} 
~\theta_k~.
\eeq
It is easy to verify that $\theta_{phys}$ is invariant under
(\ref{thetatransform}) only for even $N$.  By appropriate field
redefinitions,
any choice of $\theta_k$  can be brought to
the form $\theta_k = (-1)^{k +1} \theta_{phys}$  for some 
$\theta_{phys}$.
It  follows that for $N$ even, plugging  $\Lambda_k^4 = e^{i (-1)^{k+1}
\theta_{phys}} \Lambda^4$ and the vevs (\ref{qvevs}) into
(\ref{brokeninstantonscurve}),
the contribution of partially broken instantons is proportional to 
$\sum_{k
= 1}^{N/2} e^{i 4 \pi k \over N/2}  = 0$ (for $N > 4$).
  Thus, along the flat direction (\ref{qvevs}), the contributions of
instantons in the partially broken gauge groups cancel.

The brane picture   suggests
that the worldvolume theory becomes ${\cal N}=2$ in the infrared (i.e.
large $v$, at least for fixed $N$); at large $v$ the branes are far away
from the orbifold fixed point and
thus do not ``feel" the reduced supersymmetry.
This leads to the hope that more nonperturbative quantities could be
matched
between the deconstructed and continuum theory than just the agreement of
  $\tau$ parameters considered in this paper. We leave this for future
study.

It is also worth commenting that it may be more natural to relate
the continuum theory to the
deconstructed theory along this special flat direction,
despite the fact that the modulus that appears in the Seiberg-Witten 
curve
does not receive broken instanton corrections in either case.  Other
operators might still receive such corrections, and the nonperturbative
matching of those operators between the deconstructed and continuum
theories
may be nontrivial.
For example, along generic flat directions in the
deconstructed theory operators like $\cosh(\pi A R)$, which are related 
to the
operator $T$ in the large $N$ limit,
  are expected to receive non-perturbative correction due to
the dynamics in partially broken gauge groups.
On the other hand, it is natural to conjecture
that along the special flat direction considered in this section
all such corrections vanish.

\subsection{Large radius limit}
\label{largeradius}

The exact result for the curve should reproduce correctly the infrared
behaviour in the large-$R$
limit. The 5D $SU(2)$ theory has been studied in \cite{Seiberg:1996bd}; 
for
analysis of general 5D theories see
\cite{Intriligator:1997pq}.  In the 5D uncompactified case, the
nonrenormalization theorem
restricts the prepotential to contain at most cubic terms.
The coefficient of the cubic term is
related to the coefficient of the Chern-Simons term. In the $SU(2)$
theory that we are considering, a tree-level Chern-Simons term is not
allowed; the only
contribution to the CS coefficient occurs at one loop along the Coulomb 
branch and is
computed in \cite{Witten:1996qb}.  We will
check, in what follows, that the curve (\ref{curve5d}) reproduces these
results in the large-$R$ limit.

To begin, consider the perturbative part of the the $\tau$-parameter in 
the
deconstructed theory.
  It is clear  from  the expression in (\ref{tau1i1i}) that the
instanton contributions vanish in the $R \rightarrow \infty$ limit (the
instantons, which
are Euclidean particles in 5D,  have infinite action in this limit and so
can not contribute
to the path integral), hence  the perturbative part of $\tau$ (in the 
$\overline{DR}$-scheme) is:
\begin{equation}
\label{taupert2}
{\tau_{pert} \over 4 \pi i } =   {1 \over 4 \pi^2} \log { 4  v^{2 N} 
\sinh^2
\pi A R \over  \Lambda^{2N}} ~.
\end{equation}
Let us make some comments on the meaning of $\tau_{pert}(A)$. Using the
product formula
$\sinh x = x \prod_{n>0}(1 + x^2/(n^2 \pi^2))$, we can rearrange equation
(\ref{taupert2}) as
follows:
\beq
\label{taupert3}
{\tau_{pert} \over 4 \pi i} = {1 \over 4 \pi^2} \log {A^2 \over
   \Lambda_D^2} + {1\over 4 \pi^2}
\sum\limits_{n \ne 0}  \log \left(A^2 + {n^2 \over R^2}\right) - \log{
n^2\over R^2}  ~.
\eeq
The formula (\ref{taupert3}) has a simple physical
interpretation. It gives the perturbative  running of the  diagonal 
$SU(2)$
gauge coupling as a function of the scale $A$; recall    that ${\rm Im  }
\tau_{pert}(A) \sim
1/g_D^2(A)$. The leading  $\sim \log A$ term  accounts for the running of
the 4D coupling
at small scales $A$, obeying  $\Lambda_D \ll A \ll 1/R$. The
sum over $n \ne 0$ correctly (i.e.,  consistent with the symmetries)
   takes into account the contributions of the KK modes to the running. To
see this,
note that for fixed $A$, the main contribution to the sum in
(\ref{taupert3})
comes from modes $n \le A R$, while  the
contribution of KK modes with $n \gg A R$ cancels between the two terms 
in
the sum. Hence,
modes
of mass greater than $A$ decouple from the running of the  Wilsonian
coupling, consistent
with our interpretation of $\tau_{pert}(A)$.

Next, we can also consider the limit of large $R$ and fixed $A$. In this
limit, as
discussed in the beginning of this section, only the linear
term in $A$ (corresponding to a trilinear prepotential) survives in 
$\tau$:

\begin{equation}
\label{taupert5d}
{\tau_{pert} \over 4 \pi i }\bigg\vert_{large-R}
\rightarrow {1 \over 4 \pi^2}\left( 2 \pi R A  - \log   \left(
{\Lambda\over v }\right)^{2N} \right) ~.
  \end{equation}
Using  the definition of $\Lambda$ from eqn.~(\ref{scales}):
\beq
\Lambda^4 = 16  ~v ^4 ~\exp\left(- {8\pi^2\over g^2 (2 v)} \right) = v^4
~\exp\left(- {8\pi^2\over g^2 (v)}\right) ,
\eeq
we then obtain, at large $N$:
\beq
\label{perttau2}
{\tau_{pert} \over 4 \pi i } =2 \pi R \left( { N \over 2 \pi R  ~g^2 
(v) }
+  {A \over 4 \pi^2} \right) =
2 \pi R \left(  {1 \over g_5^2} +  {A \over 4 \pi^2} \right) ~.
\eeq
The interpretation of the two terms in (\ref{perttau2}) is as follows. 
The
overall $2 \pi R$ factor can be interpreted as an integration over the
extra dimension and the
   (dimensionful) combination $ {2 \pi R\over N} g^2(v) = v^{-1} g^2(v) =
g_5^2$   as the
5D gauge coupling. The real part of the term  linear in $A$ gives the
power-law running of the
coupling \cite{Dienes:1998vh} (recall that in the ``Weyl wedge" of the 5D
theory  Re $A > 0$ \cite{Seiberg:1996bd}).
The  imaginary part of the second term  originates in the one-loop 5D
Chern-Simons term  mentioned above.
The imaginary part of $R$ in eqn.~(\ref{perttau2}) can be made to vanish
by choosing $v$ real or,   as already mentioned in Section
\ref{pertmass},   be
interpreted as an expectation value of a field in the background
supergravity multiplet.

\section{Conclusions}
We have considered non-perturbative effects in theories with extra
dimensions
from several different perspectives: exact results, explicit instanton
calculations and deconstructed extra dimensions. For definiteness
we have focused on the 5D $SU(2)$ theory
with eight supercharges. We have shown how to perform an explicit 
one-instanton calculation in this theory by using two towers of instanton
solutions obtained from large gauge transformations acting
on the ordinary 4D instanton. Our results are  in agreement
with an improved version of the exact results obtained for this model
in \cite{Nekrasov}. In the second part of the paper, we have considered 
the
deconstructed version of the same theory. We have shown that the
Seiberg-Witten curve for the deconstructed model is in agreement with
exact results and an explicit instanton calculation for the continuum
theory, thus providing
the first non-perturbative evidence in favor of deconstruction.

\section*{Acknowledgements}
We thank Keith Dienes, Ken Intriligator, Amanda Peet, and John Terning
for useful discussions, and the Aspen Center for
Physics for its hospitality while this work was initiated.
C.C. is an Oppenheimer Fellow at the Los Alamos National Laboratory
and supported in part by a DOE OJI grant.
C.C. and J.E. are supported in part by the US Department of Energy
under contract W-7405-ENG-36. 
V.V.K.  is supported in part by a PPARC Special Project Grant.
E.P. is supported by the University of
Toronto and the Connaught Foundation.
Y. Shadmi holds a Technion Management Career Development Chair, and
thanks the Weizmann Institute for hospitality while part of this
work was completed.
Y. Shirman is supported in part by DOE grant DE-FG03-92-ER-40701.

\end{document}